\documentclass[a4paper]{article}
\usepackage{spconf,amsmath,amssymb,graphicx,multirow}
\usepackage{booktabs}

\title{Multi-query Multi-head Attention Pooling and inter-topK penalty for speaker verification}
\name{Miao Zhao$^1$, Yufeng Ma$^1$, Yiwei Ding$^{1,2}$, Yu Zheng$^1$, Min Liu$^1$, Minqiang Xu$^1{}^\dag$\thanks{\textsuperscript{\dag} Corresponding author.}}
\address{$^1$ SpeakIn Technologies Co. Ltd. \\ $^2$ Fudan University}

\ninept
\begin{document}
%
\maketitle
\begin{abstract}
This paper describes the multi-query multi-head attention (MQMHA) pooling and inter-topK penalty methods which were first proposed  in our submitted system description for VoxCeleb speaker recognition challenge (VoxSRC) 2021. Most multi-head attention pooling mechanisms either attend to the whole feature through multiple heads or attend to several split parts of the whole feature. Our proposed MQMHA combines both these two mechanisms and gain more diversified information. The margin-based softmax loss functions are commonly adopted to obtain discriminative speaker representations. To further enhance the inter-class discriminability, we propose a method that adds an extra inter-topK penalty on some confused speakers. By adopting both the MQMHA and inter-topK penalty, we achieved state-of-the-art performance in all of the public VoxCeleb test sets.
\end{abstract}

%
\section{INTRODUCTION}
Speaker verification focuses on whether two utterances come from the same speaker. Over the years, various embedding-based models have been developed to encode the utterances to an embedding with speakers' feature. There are two main categories of embedding-based models. One is the conventional i-vector which estimates the speaker representation based on Gaussian mixture model (GMM). The other is built on a deep neural network (DNN), such as the typical x-vector \cite{snyder2017deep}.

Recently, the x-vector framework has proved its superiority over the i-vector in many speaker verification tasks \cite{snyder2018x}. After that, more and more optimizations were proposed based on the original x-vector architecture. For example, ResNet \cite{he2016deep} was used to replace the time delay neural network (TDNN) layers of x-vector \cite{cai2018novel, zeinali2019but} due to its strong ability to extract the feature. ECAPA-TDNN \cite{desplanques2020ecapa} was also proposed as an enhanced version of x-vector and achieved a competitive performance with ResNet \cite{thienpondt2020idlab}. Moreover, ECAPA-TDNN was further enhanced by involving both 1D and 2D convolution neural networks (CNN), named ECAPA CNN-TDNN \cite{thienpondt2021integrating}. In general, these DNN-based architectures have four parts: (1) a backbone to encode the acoustic features to a high-level representation, (2) a pooling layer to map a variable sequence to a fixed-length embedding, (3) several segment level layers to decode the hidden information and (4) a loss function to classify different speakers and to learn a discriminative speaker embedding. In this paper, we mainly focus on the pooling layer and loss function to further enhance the performance of speaker verification.

For the DNN-based architecture, the pooling layer is a key component to aggregate the variable sequence to an utterance level embedding. Recently, the statistics pooling \cite{snyder2017deep} has been popular to represent the speaker characteristics even if there are other alternatives, such as higher-order statistics \cite{wang2021revisiting} and channel-wise correlation matrix \cite{stafylakis2021speaker}. Considering the different importance of different frames of a sequence, many works focus on the weights of frames to obtain a better segment level representation. For example, inspired by i-vector, learnable dictionary encoding (LDE) pooling \cite{cai2018novel} and recent Xi-vector \cite{lee2021xi} learned the weights based on the theory of Gaussian mixture model (GMM). Meanwhile, the simpler and more efficient self-attention mechanism were also introduced to calculate weighted mean and standard deviation, named attentive statistics (AS) pooling \cite{okabe2018attentive}. Moreover, some multi-head mechanisms were further used to increase the diversity of attention, such as self-attentive (SA) pooling \cite{zhu2018self} and self multi-head attention (MHA) pooling \cite{india2019self}. However, the two typical multi-head attention pooling have a completely different definition on heads. The SA defines the multi-head as adding more than one group of trainable parameters and the attentive weights for every head will be computed by the whole features (we prefer to name the head as a query in this case), while the MHA firstly splits the channels of features into several groups and then assigns an attentive head for each group respectively. Comparing with the SA, the MHA makes it possible to learn weights with a part of features. However, one single head in each group may be insufficient to capture the patterns of speaker characteristics. To address this issue, we proposed MQMHA pooling by adding more than one query for each group. Furthermore, inspired by channel-dependent attentive statistics (CAS) \cite{desplanques2020ecapa,thienpondt2020idlab} pooling and vector-based SA (VSA) pooling \cite{wu2020vector}, we also consider assigning a unique chann channels of one frame rather than applying the same weight on all channels. Therefore, our proposed MQMHA is a generalized pooling structure covering AS, MHA, SA, and VSA, etc.

Besides the pooling layer, the loss function is also important to learn a discriminative speaker embedding. It ensures a low similarity of different speakers and a high similarity within the same speaker. Despite the popularity of AM-Softmax \cite{wang2018additive, wang2018cosface} and AAM-Softmax \cite{deng2019arcface} in speaker verification and a great number of successes they have made \cite{liu2019large}, the same angular margin for different speakers could be inappropriate because some of the speakers are more difficult to recognize than others. Recently, setting an adaptive margin for each sample has been proposed in \cite{zhou2020dynamic}. However, with many hard samples generated in data augmentation, tuning the range of margin can be difficult. Different from this, since the relative strong penalty could be expected for similar speakers, we proposed adding an extra inter penalty to the top k negative speakers based on the original AM-Softmax loss. On one hand, the proposed inter top k penalty is different from other losses which focus on inter class, such as minimum hyper-spherical energy (MHE) \cite{liu2018learning}, in which our method focuses on the relation between a sample and its top k closed class centers but MHE pushes different centers to be uniformly distributed. Moreover, our proposed inter top k penalty could be also seen as a hard prototype mining (HPM) method without extra sampling requirements for it also pays more attention to similar speakers. Finally, by applying both MQMHA and inter-topK penalty, we achieved state-of-the-art performance in VoxCeleb tasks. 


The organization of this paper is as follows: Section 2 describes our baseline architecture based on a 34-layer ResNet. Section 3 describes two proposed methods, MQMHA and inter-topK penalty. The experiments and results are given in the Section 4. And we concludes this paper in the Section 5.

\section{Baseline System}
\begin{figure}[h]
  \centering
  \includegraphics[width=0.48\textwidth,height=0.6\textwidth]{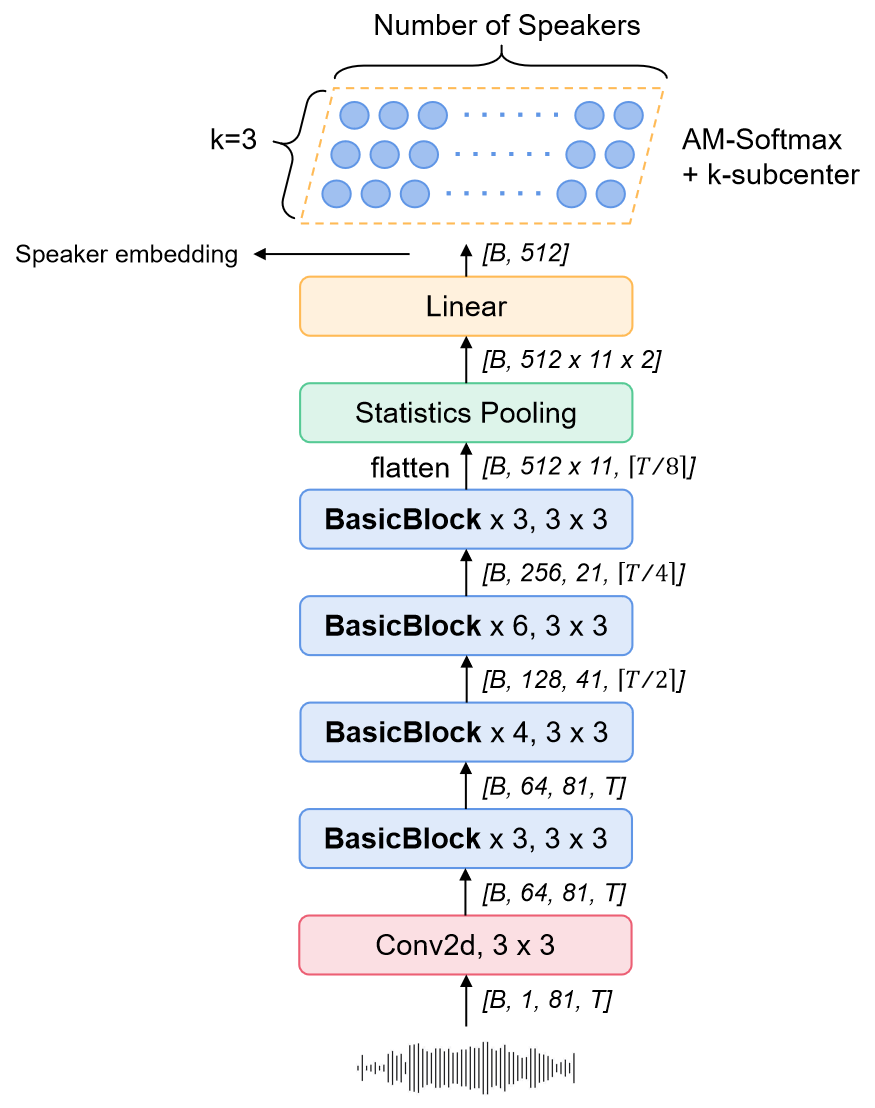}
  \caption{\textbf{Our baseline architecture for speaker verification}. The backbone is a modified 34-layer ResNet. The statistics pooling with concatenating mean and stddev is used as the basic pooling layer. The loss function is AM-Softmax with 3 sub-centers. The margin and scale of AM-Softmax are 0.2 and 35 respectively. The frequency dimension of input features is 81. \textbf{B}: the mini-batch size for training. \textbf{T}: the number of frames of input features.}
  \label{baseline}
\end{figure}

In this section, we first introduce our baseline system architecture and then describe the training protocol. As shown in \textbf{Figure \ref{baseline}}, the backbone of our baseline system is a modified version of the standard 34-layer ResNet, in which the kernel size of the first convolution is changed to 3 and the max pooling is removed. For the loss function, besides AM-Softmax, the k-subcenter \cite{deng2020sub} method is also used jointly as the basic loss function. In this case, the cosine similarity between a sample and one center of speaker is given by $cos(\theta_{i,j})=\max(||x_i||\cdot||W_{j,k}||), 1\leq k\leq K$, where the $\max$ function means that the nearest center is selected and it inhibits possible noisy samples from interfering the dominant class center.


Before training, we extracted 81-dimensional log Mel filter bank energies based on Kaldi \cite{povey2011kaldi}. The window size is 25 ms, and the frame-shift is 10 ms. 200 frames of features were extracted without extra voice activation detection (VAD), and the features were cepstral mean normalized before being fed into networks. 
During training, the SGD optimizer with a momentum of 0.9 and a weight decay of 1e-3 was used. We used 8 GPUs with 1,024 mini-batch and an initial learning rate of 0.08 to train our models. As is described above, 200 frames of each sample in one batch were adopted to avoid over-fitting and speed up training. We adopted ReduceLROnPlateau scheduler with a frequency of validating every 2,000 iterations, and the patience is 2. The minimum learning rate is 1e-6, and the decay factor is 0.1. Furthermore, the margin gradually increases from 0 to 0.2 \cite{liu2019large}. We used Pytorch \cite{paszke2019pytorch} to conduct our experiments.

After the training is done, a 512-dimensional embedding could be extracted from the linear layer and the single cosine similarity is used to compute the score of two embeddings.  

\section{The Proposed Methods}
\subsection{Multi-query Multi-head Attention Pooling}
Most attentive pooling layers pay attention to the importance of some unique features, such as giving different frames and frequency different contributions to a speaker representation. The multi-head is usually used to avoid the simple pattern learned by a single head. As described in \textbf{Section 1}, the SA and MHA give two different definitions of head and the proposed MQMHA pooling combines them to attend more patterns of the feature. Besides the definition of head, there are also two noteworthy points in various attentive pooling used in speaker verification. The first point is that most attentive poolings assign a same weight to all channels of one frame (shared case) except the recent VSA pooling. The VSA is more like the original self-attention mechanism in which every value of input features will have a unique attentive weight (unique case). The other is that the attention module of most of attentive pooling has two linear layers but MHA only has one linear layer to reduce the number of parameters. To evaluate the effects of these two points, we also combine them in our proposed MQMHA method and the final MQMHA can be described as below.

Suppose we have a backbone output $O=[o_1,o_2,...,o_T]$, $o_t\in \mathbb{R}^d$ and each $o_t$ is split into $H$ parts with $o_t=[o_{t}^{1},o_{t}^{2},...,$ $o_{t}^{H}], o_{t}^{h}\in \mathbb{R}^{d/H}$, where $H$ is the number of heads. For each head, it has $Q$ trainable queries. Then the attention weight of $w_{t,h,q}$ is defined as:
\begin{equation}
  w_{t,h,q} = \frac{\mathrm{exp}(F(o_{t}^{h}))}{\sum^T_{t=1}\mathrm{exp}\,(F(o_{t}^{h}))}
  \label{attention_weight}
\end{equation}
where the function $F(\cdot)$ is an attention mechanism to calculate weights and it can be composed of one linear layer or two linear layers with a nonlinearity:
\begin{equation}
  F(o_{t}^{h}) = 
  \begin{cases}
  (o_{t}^{h})^T w_a, &  n=1, \\ 
  relu((o_{t}^{h})^T w_b)\,w_c, & n=2.
  \end{cases}
  \label{attention_function}
\end{equation}
where $w_a$ is a matrix of size $d_h\times d_s, w_b$ is a matrix of size $d_h\times d_k$ and $w_c$ is a matrix of size $d_k\times d_s$. The $d_k$ is the hidden size of the two linear layers and it is set to 512 by default in our experiments. The $d_s$ is the number of weight and it equals to 1 for the shared case and $d_h$ for the unique case. After the weights are calculated by the attention mechanism, the representations of mean and standard deviation can be formulated as \textbf{Equation} \textbf{(\ref{mean})} and \textbf{Equation} \textbf{(\ref{std})} respectively:
\begin{equation}
  \mu_{h,q} = \sum_{i=1}^{T} w_{t,h,q} \cdot (o_{i}^{h})^T
  \label{mean}
\end{equation}
\begin{equation}
  \sigma_{h,q} = \sqrt{\sum_{i=1}^{T} w_{t,h,q} \cdot (o_{i}^{h})^T \cdot (o_{i}^{h})^T -\mu_{h,q} \cdot \mu_{h,q}}
  \label{std}
\end{equation}

Then we concatenate all of the sub-representations to get the utterance level embedding with $E_{\mu}=[\hat{\mu}_{1},\hat{\mu}_{2},...,$ $\hat{\mu}_{H}]$, where $\hat{\mu}_{h}=[\mu_{h}^{1},\mu_{h}^{2},...\mu_{h}^{Q}]$, and an extra attentive standard deviation $E_{\sigma}$ could be obtained from all of the $\sigma_{h,q}$ with the same way. Finally, this representation of standard deviation is concatenated with $E_\mu$ to enhance the performance. As mentioned above, the MQMHA contains the cases of SA ($H=1, Q=1, n=2, d_s=1$), MHA ($H>1, Q=1, n=1, d_s=1$), AS ($H=1, Q>1, n=2, d_s=1$) and VSA ($H=1, Q>1, n=2, d_s=d_h$). Moreover, learning more patterns from local features and multi queries at the same time is compatible based on the MQMHA.  

\subsection{Extra Inter-TopK Penalty for AM-Softmax Loss}
The AM-Softmax and AAM-Softmax loss functions have been widely used in recent speaker recognition works. However, the optimization to further distinguish similar speakers could be limited due to the same margin applied on all negative classes. To mitigate this issue, we proposed adding extra inter-topK penalty on AM-softmax. Given a batch with $N$ examples and a number of classes of $C$, the formulation with adding extra inter-topK penalty based on the AM-Softmax is given by:
\begin{equation}
\mathcal{L}_{AM'}\!=\!- \frac{1}{N}\!\sum_{i=1}^{N}log \frac{e^{s\cdot (cos\theta_{i,y_i}-m)}}
  {e^{s\cdot (cos\theta_{i,y_i}-m)}+\!\!\!\sum\limits_{j=1,j \neq y_i}^{C}e^{s\cdot \phi(\theta_{i,j})}}
  \label{am}
\end{equation}
where $m$ is the original margin of AM-Softmax and $s$ is the scalar of magnitude. And here the $cos\theta_{i,j}$ in AM-Softmax is already replaced by the $\phi(\theta_{i,j})$ to add an extra penalty $m'$ on inter-class:
\begin{equation}
  \phi(\theta_{i,j}) = 
  \begin{cases}
  cos\theta_{i,j}+m' &  j\in \mathop{arg\,topK}\limits_{1\leq n\leq C, n\ne y_i}(cos\theta_{i,n}) \\
  cos\theta_{i,j} & \textit{Others}.
  \end{cases}
  \label{phi}
\end{equation}
where the $cos\theta_{i,j}+m'$ could be replaced by $cos(\theta_{i,j}-m')$ for the AAM-Softmax loss function. The extra penalty $m'$ is only added for the $k$ closest centers to the example $x_i$. Since the similarity between samples and centers could be changed during training, the penalty will not always be effective for some fixed centers. Moreover, it will pay more attention to the confused pairs of different speakers as training converges and the confidence of $cos\theta_{i,j}$ increases. Therefore, this method is also a hard prototype mining method but without extra sampling requirements.

The $k$ in $top$ function is an important hyperparameter which determines the number of top nearest negative classes to select for each sample. To give an analysis of this variable, we firstly transform the \textbf{Equation (\ref{am})} to a form with inter-class penalty only where the numerator and denominator are multiplied by $e^{s\cdot m}$ at the same time:
\begin{equation}
\mathcal{L}_{AM'}\!=\!- \frac{1}{N}\!\sum_{i=1}^{N}log \frac{e^{s\cdot cos\theta_{i,y_i}}}
  {e^{s\cdot cos\theta_{i,y_i}}+\!\!\!\sum\limits_{j=1,j \neq y_i}^{C}e^{s\cdot (\phi(\theta_{i,j}) + m)}}
  \label{am_type2}
\end{equation}

Then with the different $k$, the averaged $\hat{m_k}$ between one example and all negative classes could be given by:
\begin{equation}
  \hat{m_k} = m+\frac{k\cdot m'}{C-1},\, k\in[0,\,C-1]
  \label{averaged_penalty}
\end{equation}
where the range of $\hat{m_k}$ in terms of different $k$ is $m\le \hat{m_k} \le m+m'$. Note that the $\hat{m_k}$ will be equal to $m$ when $k=0$ and equal to $m+m'$ when $k=C-1$. For these two cases, this equation is equal to the general AM-Softmax and there will be no special optimization for hard examples. While for the other cases, there are two discriminative penalties for different negative speakers. In general, the k is not expected to be too larger for the negative speakers with high similarity are usually in the minority that it is similar to the selection of imposters in adaptive score normalization.
\begin{table*}[t]
  \centering
  \caption{\textbf{Results of Pooling, Loss and Their Combination on Five Test Sets of VoxCeleb.}}
  \setlength{\tabcolsep}{0.6mm}{
  \begin{tabular}{clcccccccccc}
    \toprule
    \multirow{4}{*}{\textbf{}} &
    \multirow{4}{*}{\textbf{Method}} & 
    \multicolumn{2}{c}{\multirow{2}{*}{\textbf{VoxCeleb1-O}}} & 
    \multicolumn{2}{c}{\multirow{2}{*}{\textbf{VoxCeleb1-E}}} &  \multicolumn{2}{c}{\multirow{2}{*}{\textbf{VoxCeleb1-H}}} & 
    \multicolumn{2}{c}{\multirow{2}{*}{\textbf{VoxSRC20-dev}}} &  \multicolumn{2}{c}{\multirow{2}{*}{\textbf{VoxSRC21-val}}}   \\ \\
    \cline{3-11} 
     & & \multirow{2}{*}[-2pt]{\textbf{EER(\%)}} & \multirow{2}{*}[-2pt]{\textbf{$\textbf{DCF}_\textbf{0.01}$}} & \multirow{2}{*}[-2pt]{\textbf{EER(\%)}} & \multirow{2}{*}[-2pt]{\textbf{$\textbf{DCF}_\textbf{0.01}$}} & 
     \multirow{2}{*}[-2pt]{\textbf{EER(\%)}} & \multirow{2}{*}[-2pt]{\textbf{$\textbf{DCF}_\textbf{0.01}$}} & 
     \multirow{2}{*}[-2pt]{\textbf{EER(\%)}} & \multirow{2}{*}[-2pt]{\textbf{$\textbf{DCF}_\textbf{0.05}$}} &
     \multirow{2}{*}[-2pt]{\textbf{EER(\%)}} & \multirow{2}{*}[-2pt]{\textbf{$\textbf{DCF}_\textbf{0.05}$}} \\ \\
    \midrule
    \textbf{Baseline} & Statistics \& AM (m=0.2) & 1.0101 & 0.0997 &	1.0435 & 0.0962	&1.7668 & 0.1531 & 2.7075 & 0.1380 &2.9167&  0.1576 \\
    \midrule
    \multirow{5}{*}{\textbf{Pooling}}
    & AS (q=1, h=1) \cite{okabe2018attentive} & 1.0313 & 0.0829&	1.0224&  0.0940	&1.7356&  0.1527&	2.6863 & 0.1380&	2.9317&  0.1613 \\
    & SA (q=2, h=1) \cite{zhu2018self} & 0.9968 &  0.0800& 	1.0217  & 0.0924& 	1.7402 &  0.1493& 2.6506 & 0.1339 & 2.9233 &  0.1572 \\
    & VSA (q=2, h=1) \cite{wu2020vector} & 0.9995 & 0.0845&	1.0294&  0.0924	& 1.7483 & 0.1479&	2.6783 & 0.1333	&2.8983 & 0.1566 \\
    & MHA (q=1, h=16) \cite{india2019self} & 0.9756 & 0.0840 &	1.0270 &0.0930 &	\textbf{1.7020} & 0.1467 &	2.6450 &0.1321 &	2.7850 & 0.1503\\
    & \textbf{MQMHA (q=4, h=16)} & \textbf{0.9465} & \textbf{0.0783} &	\textbf{1.0090} & \textbf{0.0913}	& 1.7099 & \textbf{0.1465} &	\textbf{2.6172} & \textbf{0.1316}	 & \textbf{2.7467} & \textbf{0.1480} \\
    \midrule
    \textbf{Loss} & \textbf{Inter-TopK ($\textbf{m}'_{\textbf{top5}}$=0.06)} & 0.9783 &0.0846&	1.0088  &0.0883	&1.7060  &0.1461&	2.5998  &0.1317&	2.7117&  0.1491\\
    \midrule
    \multirow{2}{*}{\textbf{Combine}} & MHA \& Inter-TopK & 0.9730 & 0.0912   & 1.0170 & 0.0892   & 1.6860 & 0.1415   & 2.5760 & 0.1297   & 2.5800 & 0.1433 \\
    & \textbf{MQMHA \& Inter-TopK} & \textbf{0.9305} & \textbf{0.0738} &	\textbf{0.9809} & \textbf{0.0879} &	\textbf{1.6020} & \textbf{0.1373} &	\textbf{2.5070} & \textbf{0.1246} &	\textbf{2.5100} & \textbf{0.1403} \\
    \bottomrule
  \end{tabular}}
  \label{total_result}
\end{table*}

\section{EXPERIMENTS AND RESULTS}
\subsection{Training and Test Sets}
The VoxCele2-dev dataset \cite{chung2018voxceleb2} was used as our training set. It contains 1,092,009 utterances and 5,994 speakers in total. As the data augmentation make the system more robust, we here adopted a 3-fold speed augmentation \cite{yamamoto2019speaker, wang2020dku} at first to generate extra twice speakers. Each speech segment in this dataset was perturbed by 0.9 or 1.1 factor based on the SoX speed function. Then we obtained 3,276,027 utterances and 17,982 speakers. Then we also used RIRs \cite{ko2017study} and MUSAN \cite{snyder2015musan} to create extra four augmented copies of the original training utterances. And the data augmentation process was based on the Kaldi VoxCeleb recipe of sre16/v2. After the two augmentations, 16,380,135 utterances were generated to extract acoustic features. To evaluate our proposed methods, we used five public VoxCeleb tasks, VoxCeleb1-O, VoxCeleb1-E, VoxCeleb1-H, VoxSRC20-dev and VoxSRC21-val, which were also adopted in our system description \cite{zhao2021speakin} for the VoxSRC 2021. It is worth mentioning that the VoxSRC20-dev and VoxSRC21-val are much harder to recognize as the VoxSRC20-dev contains some out-of-domain utterances and the VoxSRC21-val focuses on multi-lingual verification.



\subsection{Results on Voxceleb Test Sets}
In our experiments, the performance is evaluated using the Equal Error Rate (EER) and the minimum Decision Cost Function (DCF) with $C_{FA}=1$, $C_M=1$, and $P_{target}=0.01$ or $P_{target}=0.05$ in different cases. \textbf{Table \ref{total_result}} shows the performance of various pooling, inter-topK loss and their combination on five test sets. For convenience, we took the performance of VoxSRC21-val as our benchmark. Firstly, our proposed MQMHA pooling outperformed all the other pooling systems, showing 5.83\% and 6.09\% relative improvement compared with the baseline in terms of EER and DCF respectively. Secondly, introducing inter-topK penalty into AM-Softmax loss reduced the EER and the DCF in all test sets, especially in VoxSRC21-val where there are more similar utterances. The EER and the DCF decreased by 7.03\% and 5.39\% respectively in comparison with the baseline. Finally, although MHA pooling and MQMHA pooling are close in performance when applied alone, when combined with inter-topK loss, MQMHA pooling achieved a better result than MHA pooling, outperforming the baseline by 13.94\% in EER and 10.98\% in DCF.

\subsection{Ablation Study of MQMHA Pooling}
To evaluate the effect of head and query in the MQMHA, we conducted an ablation study based on the VoxSRC21-val. As shown in \textbf{Table \ref{mqmha}}, a general attentive pooling (q=1, h=1, n=1, $d_{s}$=1) barely improves the performance compared to the baseline. When we start to increase the number of head, there is no obvious improvement until $h=8$ and the best results is obtained when $h=16$. With the number of heads continues to increase, the performance begins to decay. It means that the features cannot be divided to too many parts. For multi-query, the results are unstable when increasing the number of queries and keep the head as 1. However, the improvements of multi-query are significant when features are split into several parts. As for the function to calculate the attention weight, we do not observe better results when using two linear layers.

\begin{table}[h]
  \centering
  \caption{\textbf{Results of MQMHA on VoxSRC21-val.}}
  \setlength{\tabcolsep}{4mm}{
  \begin{tabular}{lcc}
    \toprule
    \textbf{Configures} & \textbf{EER(\%)} & \textbf{$\textbf{DCF}_\textbf{0.05}$} \\
    \midrule
    no attention (baseline) & 2.9167&  0.1576 \\
    q=1, h=1, n=1, $d_s$=1 &  2.8850 & 0.1569 \\
    \midrule
    q=1, h=2, n=1, $d_s$=1 & 2.9983 & 0.1717 \\
    q=1, h=4, n=1, $d_s$=1 & 2.9217 & 0.1633\\
    q=1, h=8, n=1, $d_s$=1 & 2.8200  & 0.1573\\
    q=1, h=16, n=1, $d_s$=1 & \textbf{2.7850} & \textbf{0.1503}\\
    q=1, h=32, n=1, $d_s$=1 & 2.9167 & 0.1585\\
    \midrule
    q=2, h=1, n=1, $d_s$=1 & \textbf{2.8717} & \textbf{0.1575}\\
    q=4, h=1, n=1, $d_s$=1 & 2.9233 & 0.1645\\
    q=8, h=1, n=1, $d_s$=1 & 2.8983 & 0.1642\\
    \midrule
    q=2, h=16, n=1, $d_s$=1 & 2.8367 & 0.1581\\
    \textbf{q=4, h=16, n=1, $\textbf{\emph{d}}_\textbf{s}$=1} & \textbf{2.7467} & \textbf{0.1480} \\
    q=8, h=16, n=1, $d_s$=1 & 2.7767 & 0.1557\\
    \midrule
    q=4, h=16, n=2, $d_s$=1 & 2.7800 & 0.1532 \\
    q=4, h=16, n=2, $d_s$=$d_h$ & 2.8867 & 0.1551\\
    \bottomrule
  \end{tabular} 
  \label{mqmha}}
\end{table}

\subsection{Ablation Study of Inter-topk Penalty}
For the inter-topK method, both the extra inter margin penalty $m'$ and the number of top nearest negative speakers $k$ have an important effect on the performance. As shown in \textbf{Table \ref{inter_topk}}, our proposed inter-topK outperforms baseline by 7.03\% in EER and 5.39\% in DCF when $m'=0.06$ and $k_{top}=5$. Firstly, the $k_{top}$ is a more important hyperparameter than $m'$. As described in \textbf{Equation (\ref{averaged_penalty})}, it will be the general AM-Softmax case if the $k$ is equal $0$ or $C-1$. However, simply increasing the margin from 0.20 to 0.26 does not improve the speaker verification performance. On the other hand, only adding extra penalty term $m'$ on those top $k$ negative classes significantly improves the system performance. The best result is obtained when $k$ equals 5. We also observe that $k$ should not be too large e.g. $k=10$ due to the fact that some negative classes may be overly punished. Similarly, the extra penalty $m'$ neither should not be too large.

\begin{table}[h]
  \centering
  \caption{\textbf{Results of Inter-TopK on VoxSRC21-val}.}
  \setlength{\tabcolsep}{3.0mm}{
  \begin{tabular}{lcc}
    \toprule
    \textbf{Configures} & \textbf{EER(\%)} & \textbf{$\textbf{DCF}_\textbf{0.05}$} \\
    \midrule
    m=0.20, $m'$=0.00 (baseline) & 2.9167 &  0.1576 \\
    m=0.22, $m'$=0.00 & 2.9133 & 0.1656 \\
    m=0.24, $m'$=0.00 & 2.9200 & 0.1616 \\
    m=0.26, $m'$=0.00 & 3.0000 & 0.1719 \\
    \midrule
    m=0.20, $m'$=0.02 \& $k_{top}$=5 & 2.7550 & 0.1605 \\
    m=0.20, $m'$=0.04 \& $k_{top}$=5 & 2.7450 & 0.1506 \\
    \textbf{m=0.20, $\textbf{\emph{m}}'$=0.06 \& $\textbf{\emph{k}}_\textbf{\emph{top}}$=5} & \textbf{2.7117} & \textbf{0.1491} \\
    m=0.20, $m'$=0.08 \& $k_{top}$=5 & 2.7233 & 0.1501 \\
    \midrule
    m=0.20, $m'$=0.06 \& $k_{top}$=1 & 2.8833 & 0.1605 \\
    m=0.20, $m'$=0.06 \& $k_{top}$=2 & 2.7633 & 0.1543 \\
    \textbf{m=0.20, $\textbf{\emph{m}}'$=0.06 \& $\textbf{\emph{k}}_\textbf{\emph{top}}$=5} & \textbf{2.7117} & \textbf{0.1491} \\
    m=0.20, $m'$=0.06 \& $k_{top}$=10 & 2.7617 & 0.1556 \\
    \bottomrule
  \end{tabular} 
  \label{inter_topk}}
\end{table}

\section{CONCLUSION}
In this papaer, we proposed two methods, MQMHA pooling and inter-topK penalty based on AM-Softmax loss function, to further improve the performance of speaker verification. The MQMHA calculates the weights of frames by not only splitting the features to several parts along the channel axis but also assigning more than one queries for each part. The inter-topK penalty further enhances the inter class discriminability through adding an extra penalty term on top $k$ negative speakers. Both these two methods outperform our baseline model. With a combination of the two methods above, our system achieves state-of-the-art performance. The EER on VoxCeleb1-H is 1.6020\% and the corresponding minDCF is 0.1373.

\vfill\pagebreak
\bibliographystyle{IEEEbib}
\bibliography{mybib}
\end{document}